\documentclass[twocolumn,letterpaper,nofootinbib]{revtex4-1}
\usepackage{graphics}
\usepackage{graphicx}
\usepackage{epsfig}
\usepackage{units}
\begin{document}

\title{Photon operators for lattice gauge theory}

\author{Randy Lewis}
\affiliation{Department of Physics and Astronomy, York University,
Toronto, Ontario, Canada M3J 1P3}

\author{R. M. Woloshyn}
\affiliation{TRIUMF, 4004 Wesbrook Mall, Vancouver,
British Columbia, Canada V6T 2A3}

\begin{abstract}
Photon operators with the proper $J^{PC}$ quantum numbers are constructed, including one made of elementary plaquettes.
In compact U(1) lattice gauge theory, these explicit photon operators are shown to permit direct confirmation of the massive and massless states on each side of the phase transition.
In the abelian Higgs model, these explicit photon operators avoid some excited state contamination seen with the traditional composite operator, and allow more detailed future studies of the Higgs mechanism.
\end{abstract}

\maketitle

\section{Introduction}
\label{intro}

In a pioneering work, Berg and Panagiotakopoulos \cite{Berg:1983is} showed the existence
of a massless photon in compact U(1) lattice gauge theory. The massless
state that they found corresponds, in the continuum, to an axial vector
superposition of two opposite-helicity nonzero-momentum photons. The
operator used to interpolate this state transforms as $T_{1}^{+-}$ under
lattice rotations, parity transformation and charge conjugation.
($T_1$ is the cubic lattice representation for angular momentum $J=1$.)
This operator is constructed from a combination of four plaquettes and has been
used in a number of subsequent lattice studies, for example
\cite{Lee:1986tb,Lee:1986mr,Evertz:1986ur,Majumdar:2003xm,De:2017sjt,Woloshyn:2017rhe}.
In their paper, Berg and Panagiotakopoulos \cite{Berg:1983is} mention the possibility
of using operators constructed from 8-link Wilson loops \cite{Berg:1982kp} to investigate
states that transform as $T_{1}^{--}$, which corresponds to the quantum
numbers of a single photon on the lattice. As far as we are aware,
such operators have not been implemented in a lattice simulation until
now.

In this paper we investigate a number of different operators that
can be used to interpolate the photon. In Sec.~\ref{ops} examples of operators
with $T_{1}^{--}$ transformation properties are constructed using
8-link Wilson loops. As well it is shown that, although a single plaquette
does not contain any $T_{1}^{--}$, a particular linear combination of 8
elementary plaquettes does form a pure $T_{1}^{--}$ operator.

A numerical simulation of compact U(1) lattice gauge theory is discussed
in Sec.~\ref{u1}. As is well known, this theory has a weak-coupling 
unconfined phase in which free photons should exist \cite{Guth:1979gz}. 
Correlation functions of
the operators presented in Sec.~\ref{ops} were calculated in this phase
and are shown to describe a state with a dispersion relation consistent
with a massless particle. Furthermore, it is shown that although there
are multiple $T_{1}^{--}$ operators they are all propagating the
same state.

The $T_{1}^{+-}$ four-plaquette operator of Ref.~\cite{Berg:1983is} is adequate
to expose the massless photon but there are situations where an operator
with the proper photon transformation properties is needed. As an
example we consider the lattice version of the abelian Higgs 
model \cite{Higgs:1966ev},\emph{ i.e.}
a field theory of a self-coupled charged scalar field. As a function
of its parameters this model has a Coulomb phase in which charged
particles and massless photons exist, but the model also has confined and Higgs
regions where charged and massless states disappear from the spectrum 
\cite{Fradkin:1978dv,Jansen:1985cq}.
The abelian Higgs model has been studied extensively using nonperturbative
lattice methods 
\cite{Jansen:1985cq,Jansen:1985nh,Azcoiti:1990ne,Alonso:1993tv,Baig:1998ui,Romero-Lopez:2018zyy}.
To expose the Higgs boson and the massive vector
boson expected in the Higgs regime, it is typical to use composite gauge-invariant
operators \cite{Jansen:1985nh} constructed from the scalar field and gauge field links.
Here we focus on the vector boson. On the lattice it has
transformation properties $T_{1}^{--}$ like the photon. To
demonstrate that the photon acquires a mass in the Higgs regime, and
that it mixes with and describes the same state as the composite vector
boson operator, requires a photon operator with the correct quantum
numbers. This is discussed in Sec.~\ref{abhiggs}.

Section \ref{summary} gives a summary and mentions possible future work.

\section{Photon operators}
\label{ops}

The simplest gauge-invariant operator made of gauge links is the elementary plaquette.
With three spatial planes and two orientations (clockwise and counterclockwise),
there are 6 spatial plaquettes in total.
Standard group theory methods allow a calculation of the character table,
and from that the multiplicities, resulting in
\begin{equation}
{\rm plaquette} = A_1^{++} \oplus E^{++} \oplus T_1^{+-} \,.
\end{equation}
Notice that dim($A_1$)+dim($E$)+dim($T_1$)=1+2+3=6 is the number of
plaquettes, as required.
Also notice that there is no $T_1^{--}$ in the elementary plaquette so it cannot couple to a single photon.
The 3 imaginary parts of the plaquettes give the $T_1^{+-}$ and the 3 real parts give the $A_1^{++}$ and $E^{++}$.

To construct an operator with nonzero momentum, it is convenient to replace
each individual plaquette with the sum of 4 plaquettes in the same plane that
touch a specific lattice site $x$.
This does not change the group theory given above.
We refer to the imaginary part of this 4-plaquette operator as $O_1$,
\begin{equation}
O_1^i(x) = {\rm Im}\bigg(\vcenter{\hbox{\scalebox{0.2}{\includegraphics{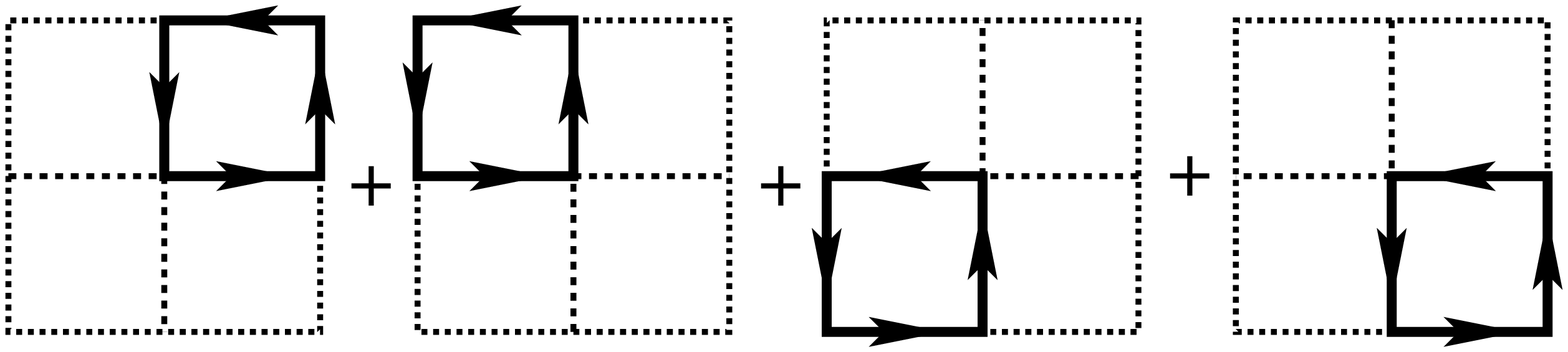}}}}\bigg)
\end{equation}
where $x$ is at the center of each diagram and $i$ is orthogonal to the page.
Previous authors
\cite{Berg:1983is,Lee:1986tb,Lee:1986mr,Evertz:1986ur,Majumdar:2003xm,Woloshyn:2017rhe}
have used this $O_1$ to couple to a pair of photons.

The same group theory approach reveals 8-link Wilson loops that do contain $T_1^{--}$.  Some examples are listed in Table 3.2 of \cite{Berg:1982kp}, including two ``figure eight'' paths that they call \#12 and \#13 with the following contents:
\begin{eqnarray}
\#12 &:& A_1^{++} \oplus A_2^{++} \oplus 2E^{++} \oplus T_1^{--} \oplus T_2^{--} \,, \\
\#13 &:& A_1^{++} \oplus E^{++} \oplus T_2^{++} \oplus T_1^{--} \oplus T_2^{--} \,.
\end{eqnarray}
We take an additional step to extract the pure $T_1^{--}$ from each, thus creating our
$O_2$ and $O_3$:
\begin{equation}
O_2^i(x) = {\rm Im}\bigg(\vcenter{\hbox{\scalebox{0.2}{\includegraphics{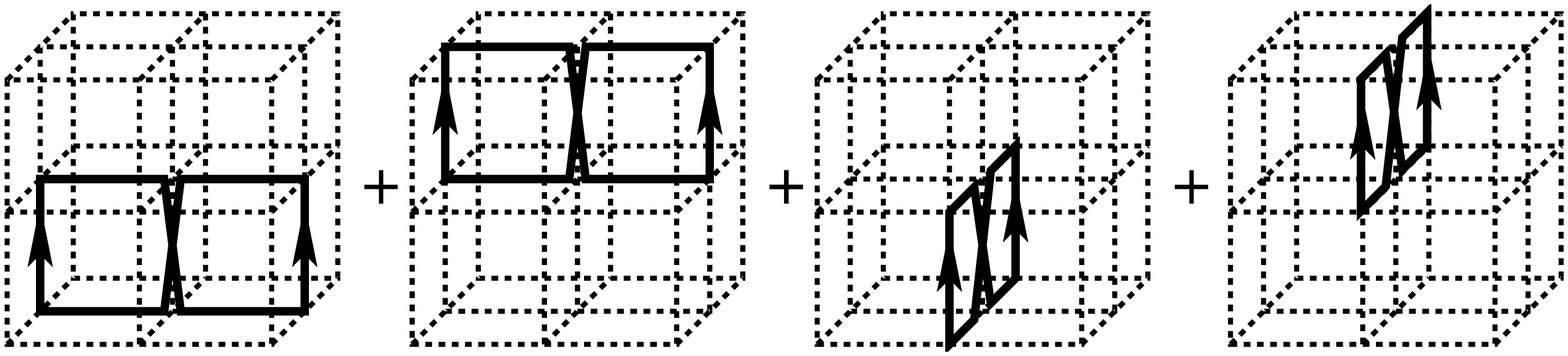}}}}\bigg) \,,
\end{equation}
\begin{equation}
O_3^i(x) = {\rm Im}\bigg(\vcenter{\hbox{\scalebox{0.2}{\includegraphics{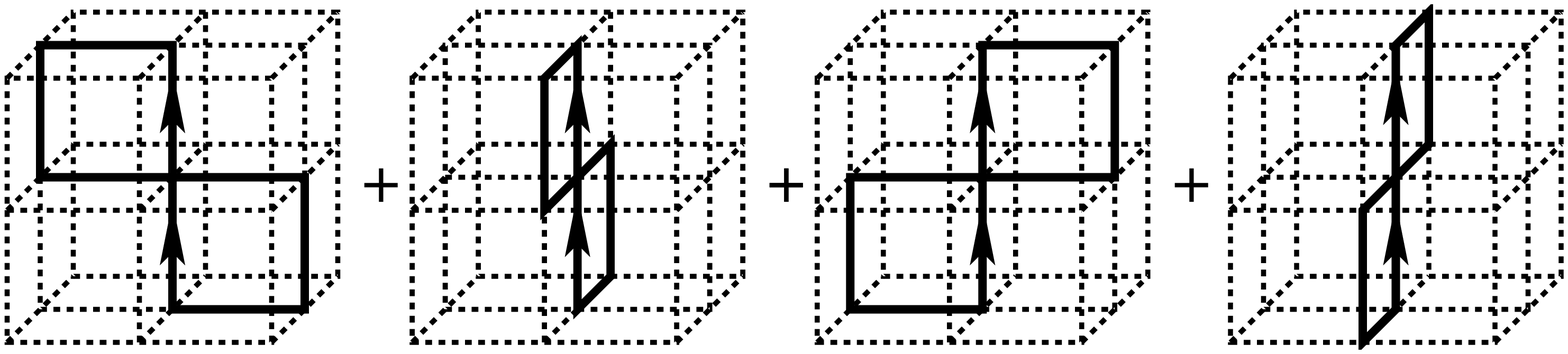}}}}\bigg)
\end{equation}
where $x$ is at the center of each diagram\footnote{Another option for $O_2^i(x)$ is to include
8 more terms in the sum, where $x$ can be at each of the 4 corners of the Wilson loop in the 2 available planes.} and $i$ points toward the top of the page.
Each of these two operators is pure $T_1^{--}$ and will couple to a single photon.

Although the elementary plaquette does not contain $T_1^{--}$, one can build a sum
of elementary plaquettes that does.
Moreover, an operator can be constructed from elementary plaquettes that couples \emph{only} to
$T_1^{--}$.
The result, denoted by $O_4$, is a sum of 8 elementary plaquettes.
For compactness, we draw all 8 plaquettes in a single diagram:
\begin{equation}
O_4^i(x) = {\rm Im}\bigg(\sum_{P=1}^8\vcenter{\hbox{\scalebox{0.2}{\includegraphics{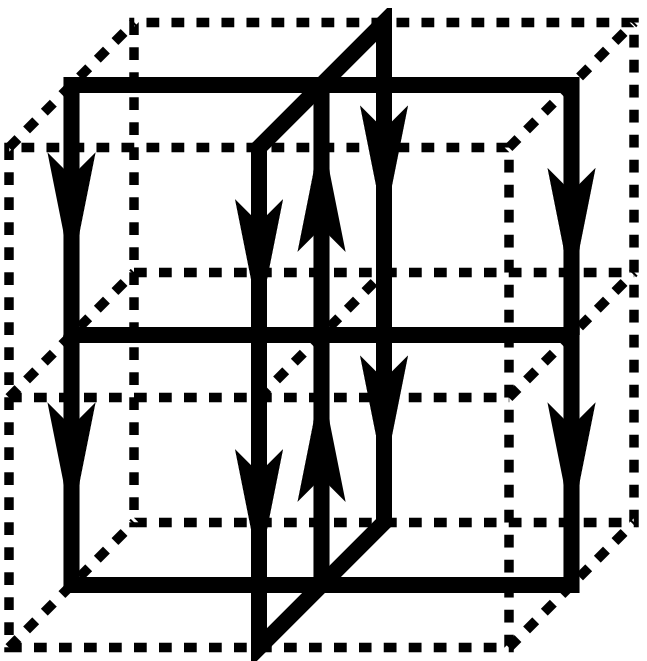}}}}\bigg) \,,
\end{equation}
where $x$ is at the center of each diagram and $i$ points toward the top of the page.
This diagram shows that the link directions are reminiscent of a solenoid.
This operator is found to produce an excellent signal for the photon.

Explicit expressions for $O_1$, $O_2$, $O_3$ and $O_4$ are given in the Appendix.
Other $T_1^{--}$ operators could be constructed as well, and Table 3.2 of \cite{Berg:1982kp}
provides a good starting point since it lists several Wilson loops containing a $T_1^{--}$
component.
We chose the two operators (\#12 and \#13) from Table 3.2
that contain no $T_1^{+-}$ component and, being planar, they are also straightforward to implement.
Likewise our $O_4$ operator is not the only sum of plaquettes that is purely $T_1^{--}$ but,
having only 8 plaquettes, it is a convenient operator and 
in numerical simulations it turns out to have the smallest statistical fluctuations among our list of operators.

\section{Compact U(1) lattice gauge theory}
\label{u1}

The compact U(1) gauge theory on a hypercubic lattice is described
by the action
\begin{equation}
\label{eq:compu1}
S_{G}=-\frac{\beta}{2}\sum_{P}(U_{P}+U_{P}^{*})
\end{equation}
 where $U_{P}$ are products of links around the elementary plaquettes.
In terms of real phase angles the gauge field links $U_{\mu}(x)$
are $e^{i\theta_{\mu}(x)}.$ At strong coupling, \emph{i.e.} small
$\beta$, the theory is confining due to the self interaction induced
by exponentiating the gauge field. At $\beta=\beta_{c}\approx$ 1.01
there is a transition to an unconfined phase \cite{Guth:1979gz,Jersak:1984pp,Arnold:2002jk}.
For $\beta>\beta_{c}$ there
should be a massless vector state (photon) in the spectrum.

\begin{figure}
\scalebox{0.45}{\includegraphics*{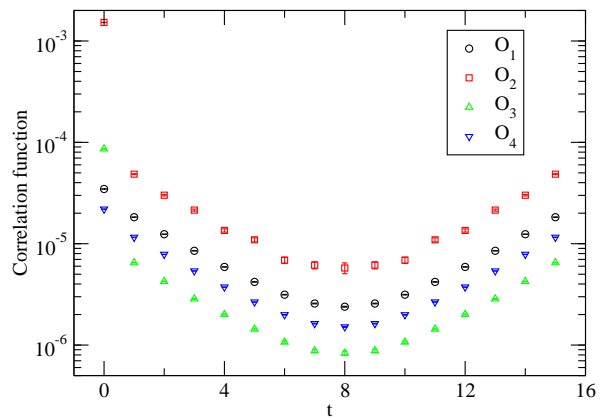}}
\caption{Correlation functions of photon operators in compact U(1) 
lattice gauge theory at $\beta$ = 1.2 plotted versus Euclidean timestep $t$.}
\label{fig:cfbeta12}
\end{figure}

\begin{figure}
\scalebox{0.45}{\includegraphics*{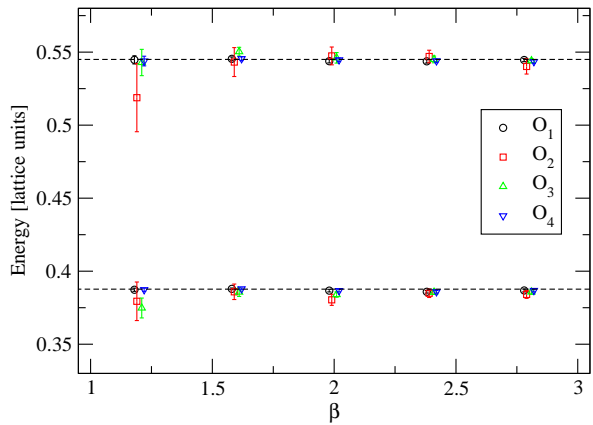}}
\caption{Energies extracted from correlators of photon operators in compact U(1) 
lattice gauge theory as a function of $\beta$. Lower points are for one unit of lattice
momentum. Upper points are for two units. The dashed lines show
energies calculated using the lattice dispersion relation for a massless particle.}
\label{fig:Evsbeta}
\end{figure}

Numerical simulations were carried out on $16^{4}$ site lattices
with periodic boundary conditions for a variety of $\beta$ values
up to $\beta=$ 2.8. A multi-hit Metropolis updating algorithm was
used. As an example of the correlation functions that were obtained,
Fig.~\ref{fig:cfbeta12} shows the diagonal correlators at $\beta$ =
1.2 of the operators discussed in Sec.~\ref{ops} projected with one unit
of lattice momentum ($\vec{p}=\frac{\pi}{8}(1,0,0)$ and permutations
with momentum transverse to the vector operator). Energies were obtained
using a constrained two-exponential fit \cite{Lepage:2001ym} over the whole time range
(excluding the source point). The results for $\beta>\beta_{c}$ are
shown in Fig.~\ref{fig:Evsbeta} for operators projected with one and two
lattice units of momentum. The dashed lines in the figure are the
energies calculated using the lattice dispersion relation
\begin{equation}
2\cosh(E)=m^{2}+8-2{\displaystyle \sum_{i}}\cos(p_{i})
\label{eq:latdr}
\end{equation}
with $m^{2}=0$.

\begin{figure}
\scalebox{0.45}{\includegraphics*{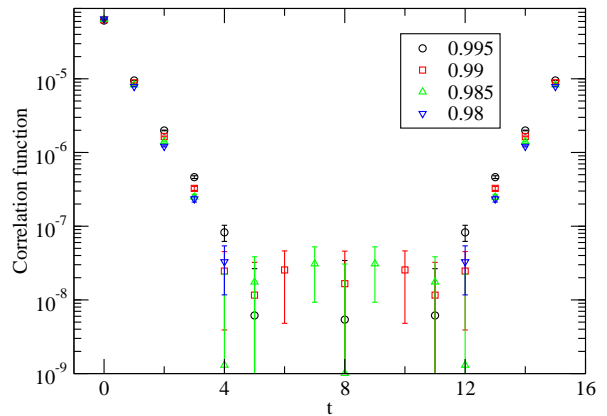}}
\caption{Momentum-projected correlation functions of $O_4$ at $\beta$ values just below $\beta_c\approx1.01$
in compact U(1) lattice gauge theory.}
\label{fig:cfO4}
\end{figure}

Below $\beta_{c}$ the photon operators propagate massive states.
This can be inferred, for example, from the momentum-projected correlation
functions of $O_{4}$ calculated at some values of $\beta$ just below
the critical value and plotted in Fig.~\ref{fig:cfO4}. The correlators
survive only a few timesteps before disappearing into noise so it
is difficult to determine the asymptotic value of the ground state
energy. However, the rate of falloff of the correlation functions
where they are statistically significant would indicate a mass greater
than one, \emph{i.e.} larger than the cutoff scale. This is consistent
with the results of Refs.~\cite{Berg:1983is,Nakamura:1991ww}.

To confirm that our set of four operators is providing evidence that the
theory contains just one photon, we can study the full 4$\times$4
correlation matrix of sources and sinks.
First, the correlation function of $O_1$ with any of the others is found to
be statistically consistent with zero at every timestep, as expected
because $O_1$ contains no $T_1^{--}$ component while the others are exclusively $T_1^{--}$.
Next we consider all elements of the 3$\times$3 correlation matrix for $O_2$, $O_3$ and $O_4$.

\begin{figure}
\scalebox{0.32}{\includegraphics*{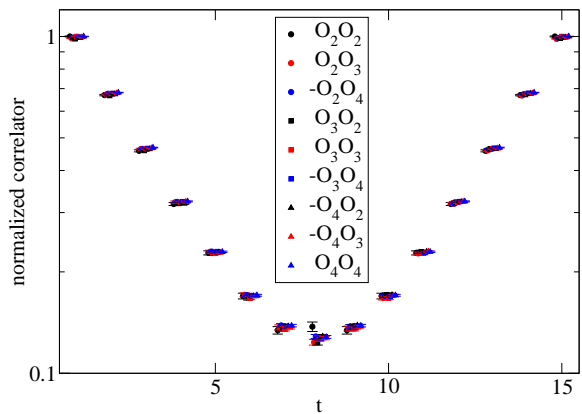}}
\caption{Entries in the 3$\times$3 correlation matrix for $O_2$, $O_3$ and $O_4$ in compact U(1) lattice gauge theory with one unit of momentum at $\beta=2$.}
\label{fig:plot3x3}
\end{figure}

The normalization of each operator is chosen such that its diagonal correlator is unity at $t=1$
(which is right beside the source).
This rescales all 9 elements of the correlation matrix as follows:
\begin{equation}
C_{ij}(t) \to \frac{C_{ij}(t)}{\sqrt{C_{ii}(1)C_{jj}(1)}}
\end{equation}
where $i\in(2,3,4)$ and $j\in(2,3,4)$.
The resulting correlation functions are shown in Fig.~\ref{fig:plot3x3}.
Data sets in this plot have a small horizontal offset for clarity, but the main point
is that all nine sets are essentially indistinguishable up to some overall minus signs.
Mathematically, the plot says our correlation matrix is proportional to
\begin{equation}
M = \left(\begin{array}{rrr} 1 & 1 & -1 \\ 1 & 1 & -1 \\ -1 & -1 & 1 \end{array}\right)
\end{equation}
which has a pair of vanishing eigenvalues and a single eigenvalue equal to 3.
The corresponding eigenvectors are
\begin{eqnarray}
\lambda_1=0 &~~\Rightarrow~~& v_1 = O_2 - O_3 \,, \\
\lambda_2=0 &~~\Rightarrow~~& v_2 = O_2 + O_3 + 2O_4 \,, \\
\lambda_3=3 &~~\Rightarrow~~& v_3 = O_2 + O_3 - O_4 \,.
\end{eqnarray}
Calculations of the $v_iv_j$ correlation functions confirm that all of them are statistically
consistent with zero at every timestep, except $v_3v_3$ which alone retains the original
photon signal from Fig.~\ref{fig:plot3x3}.
Therefore the original operators $O_i$ are all coupling to the same photon, and
$v_3$ is maximizing our overlap with that photon.

\section{Abelian Higgs model}
\label{abhiggs}

\begin{figure}
\scalebox{0.5}{\includegraphics*{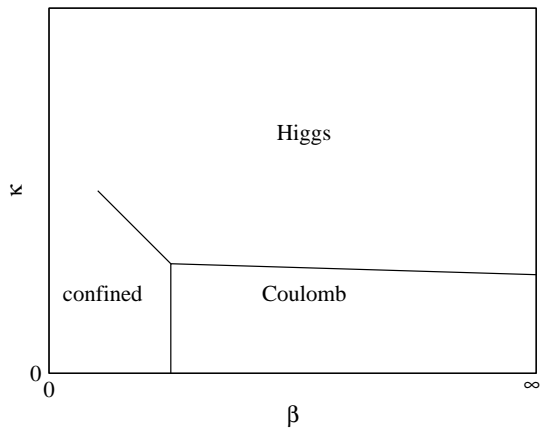}}
\caption{Phase diagram of the abelian Higgs model.}
\label{fig:phase}
\end{figure}

As an application for the operators discussed in the previous section,
we consider the abelian Higgs model. The lattice action for this model
is $S=S_{G}+S_{\varphi}$ where the gauge action was defined in Eq.~(\ref{eq:compu1})
and 
\begin{eqnarray}
\label{eq:sphi}
\nonumber
S_{\varphi} & = & -\kappa\sum_{x,\mu}(\varphi^{*}(x)U_{\mu}(x)\varphi(x+\hat{\mu})+h.c.)\\
 &  & +\sum_{x}\varphi^{*}(x)\varphi(x)+\lambda\sum_{x}(\varphi^{*}(x)\varphi(x)-1)^{2}.
\end{eqnarray}
 The phase diagram at fixed $\lambda$ is shown schematically in Fig.~\ref{fig:phase}.
Numerical calculations for this work were done assuming
that the complex scalar field $\varphi$ has a fixed unit norm,
\emph{i.e.} the value in the limit $\lambda\rightarrow\infty.$ In this
limit only the hopping term of $S_{\varphi}$ is needed and this saves
some time in doing the simulation. The interest here is in showing
the existence of a massless photon in the Coulomb phase and the fate
of the photon in the Higgs region and we don't expect these qualitative
aspects to depend on $\lambda.$ In addition to the four operators
discussed in previous sections, the composite vector boson operator
\begin{equation}
\label{eq:boson}
O_{5}=\mathrm{Im}\varphi^{*}(x)U_{i}(x)\varphi(x+\hat{i})
\end{equation}
is considered here.

\begin{figure}
\scalebox{0.45}{\includegraphics*{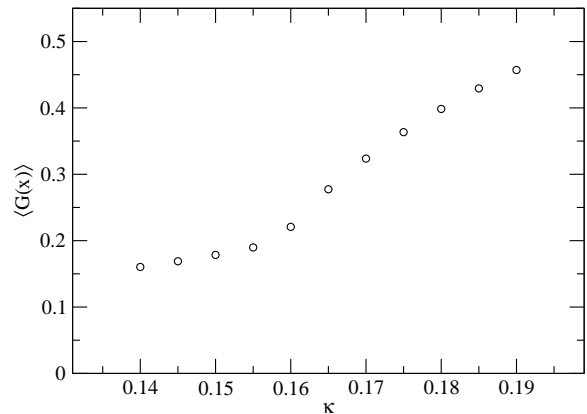}}
\caption{The expectation value of $G(x)$ as a function of $\kappa$ for the abelian Higgs
model at $\beta$ = 2 and $\lambda$ = $\infty$.}
\label{fig:phiuphi}
\end{figure}

\begin{figure}
\scalebox{0.45}{\includegraphics*{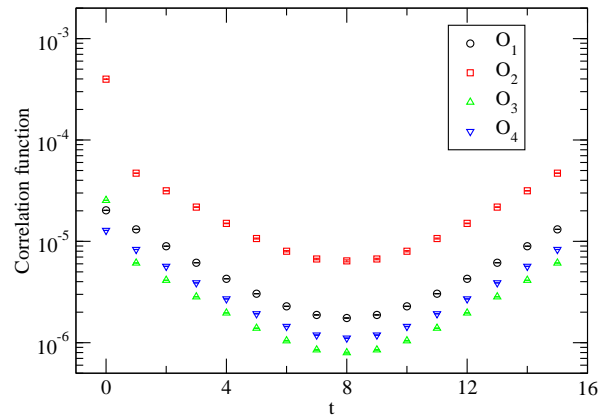}}
\caption{Correlation functions of photon operators in the abelian Higgs
model at $\kappa$ = 0.145, $\beta$ = 2 and $\lambda$ = $\infty$.}
\label{fig:cfkappa145}
\end{figure}

\begin{figure}
\scalebox{0.45}{\includegraphics*{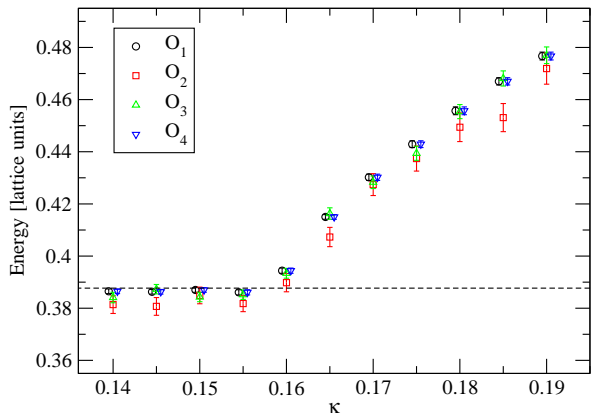}}
\caption{Energies extracted from correlators of photon operators projected with 
one unit of momentum in the abelian 
Higgs model as a function of $\kappa$ at $\beta$ = 2 and $\lambda$ = $\infty$. 
The dashed line shows energy calculated using the lattice dispersion relation 
for a massless particle.}
\label{fig:Evskappa}
\end{figure}

Numerical simulations were carried out on $16^{4}$ lattices for a
range of $\kappa$ values at fixed $\beta$ = 2, well above the critical
$\beta$ of the pure gauge theory. Although strictly speaking not
an order parameter, the quantity
\begin{equation}
\label{eq:phiuphi}
G(x)=\mathrm{Re}{\displaystyle \sum_{\mu}}\varphi^{*}(x)U_{\mu}(x)\varphi(x+\hat{\mu})
\end{equation}
is used to locate the transition region from the Coulomb to the Higgs
phase. The results are plotted in Fig.~\ref{fig:phiuphi} where one can
see that the critical $\kappa$ is around 0.16. Diagonal correlators
of the operators $O_{1}$ to $O_{4}$ projected with one unit of momentum
were calculated for a range of $\kappa$ values using a sample of
80000 field configurations. Fig.~\ref{fig:cfkappa145} showing the results
at $\kappa$ = 0.145 illustrates the quality of the correlators. The
energies extracted from diagonal photon correlators are plotted in
Fig.~\ref{fig:Evskappa}. The dashed line in the figure shows the energy
expected for a zero-mass particle. A massless photon exists in the
Coulomb region but beyond $\kappa$ = 0.16 the increasing energies
indicate an increasing nonzero mass. 

\begin{figure}
\scalebox{0.45}{\includegraphics*{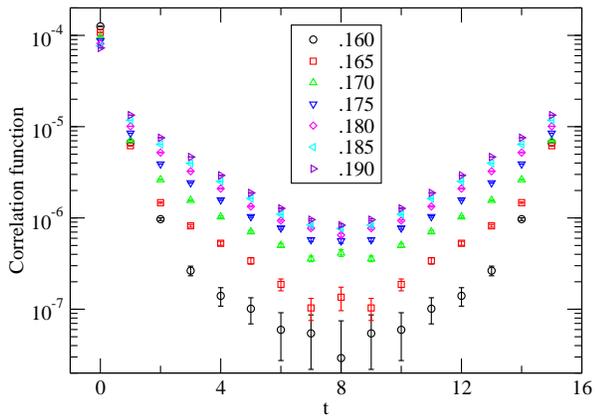}}
\caption{Correlation functions of the composite vector operator $O_5$ in the abelian Higgs
model at $\beta$ = 2 and $\lambda$ = $\infty$.}
\label{fig:cfvec}
\end{figure}

\begin{figure}
\scalebox{0.45}{\includegraphics*{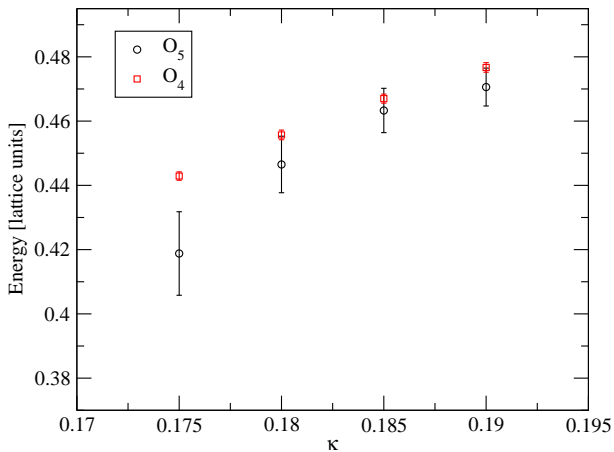}}
\caption{Energies extracted from the correlators of $O_5$ and $O_4$ in the Higgs
region of the abelian Higgs model.}
\label{fig:vecenergy}
\end{figure}

The correlation function of the composite boson operator $O_{5}$
is plotted in Fig.~\ref{fig:cfvec} for $\kappa$ values in the Higgs region.
Near the transition the correlators are too noisy to obtain a reliable
estimate of the ground state energy. At $\kappa$ = 0.175 and above,
the ground state energy can be determined reasonably well and the
values are shown in Fig.~\ref{fig:vecenergy} for correlators projected
with one unit of momentum. For comparison the energies extracted
from the correlator of $O_{4}$ are also shown.

In the semi-classical treatment of the abelian Higgs model, the massive
vector boson appears as an elementary field \cite{Higgs:1966ev} and it is natural to interpret
it as a photon having acquired a mass. In contrast, nonperturbative
lattice calculations typically use the gauge-invariant composite operator
$O_{5}$ to reveal the presence of a massive vector state \cite{Jansen:1985nh}. 
As shown
here there are photon interpolating operators which, when used in
the Higgs region, exhibit a mass compatible with that of the composite
vector boson. The question is whether the photon operators and the
operator $O_{5}$ are propagating the same state or not. To answer
this question, cross correlations between different operators are needed.
In previous studies \cite{Jansen:1985nh} where only the simple 
plaquette operator $O_{1}$
was considered, the mixing of the photon with the composite vector
could not be addressed. Recall that operator $O_{1}$ transforms as
$T_{1}^{+-}$ and has no correlation with $O_{5}$ which transforms
as $T_{1}^{--}.$ What is different in this work is that operators
with the appropriate photon transformation properties $T_{1}^{--}$
have been introduced so cross correlations can be studied.

\begin{figure}
\scalebox{0.45}{\includegraphics*{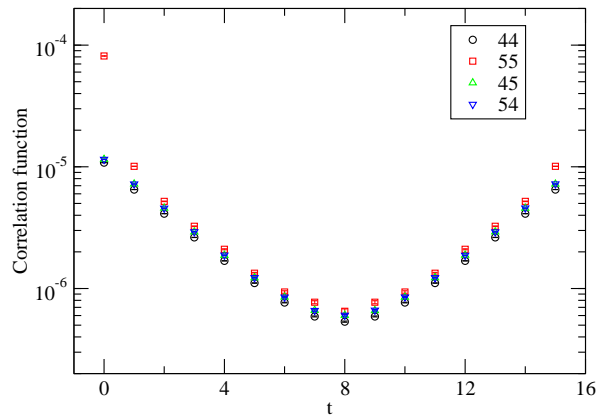}}
\caption{The four correlation functions constructed from the operators $O_4$ and  $O_5$
at $\kappa$ = 0.180.}
\label{fig:cf45180}
\end{figure}

It was shown in Sec.~\ref{u1} that the operators $O_{2},\: O_{3}$ and $O_{4}$
all propagate the same photon state so, to study cross correlations with $O_5$,
the operator $O_{4}$ which has the smallest statistical fluctuations
is chosen. Results will be shown here for $\kappa$ = 0.180. Other $\kappa$'s at
0.175 and above show similar behaviour. The four correlation functions
(momentum projected) that can be constructed using operators $O_{4}$
and $O_{5}$ are shown in Fig.~\ref{fig:cf45180}. It is apparent even without
a quantitative fit that the diagonal correlator of $O_{5}$ contains
a substantial heavy non-ground state contribution which is not propagated
by the photon operator $O_{4}.$ The cross correlators falling nicely
in between the diagonal correlators is already an indication of the
high degree of correlation between the operators.

\begin{figure}
\scalebox{0.45}{\includegraphics*{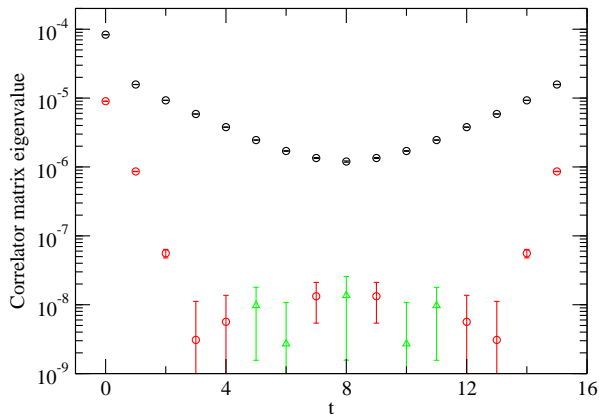}}
\caption{Eigenvalues of the 2$\times$2 correlation matrix constructed from the operators $O_4$ and  $O_5$
at $\kappa$ = 0.180. The triangles denote points where the eigenvalue is negative.}
\label{fig:ev45180}
\end{figure}

\begin{figure}
\scalebox{0.45}{\includegraphics*{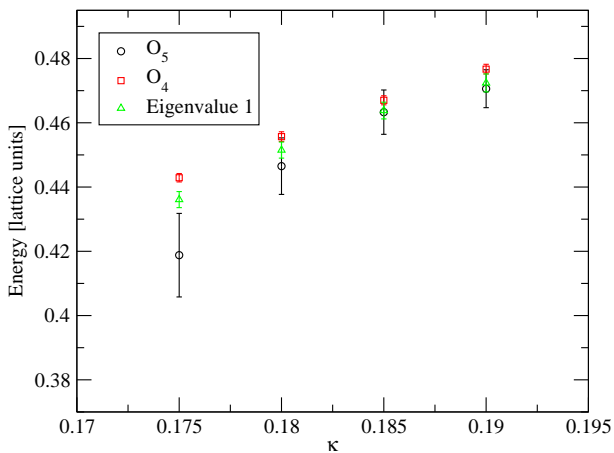}}
\caption{Energies extracted from the large eigenvalue of the 
2$\times$2 correlator matrix compared to values from the 
correlators of $O_5$ and $O_4$ in the Higgs
region of the abelian Higgs model.}
\label{fig:vecenergy2}
\end{figure}

The eigenvalues of
the 2$\times$2 correlator matrix were calculated at every timestep
and are plotted in Fig.~\ref{fig:ev45180}. The two eigenvalues are very
different in magnitude and time dependence. The ground state energy
extracted from the large eigenvalue is compatible with that extracted
from the diagonal correlators alone. This is shown in Fig.~\ref{fig:vecenergy2}
along with results for other $\kappa$ values. The small eigenvalue
is statistically significant only near the source and reflects the
fact that $O_{5}$ can excite some high lying states. This eigenvalue
is too small to measure at larger time separations which is an indication
that the operators $O_{4}$ and $O_{5}$ propagate the same ground
state. This result reminds us that in the nonperturbative context
the view that the massive vector boson present in the Higgs regime
is a massive photon is too simplistic and does not account for the
fact that in a field theory all allowable field configurations can
contribute to physical states. 

\section{Summary}
\label{summary}

Past studies of lattice theories containing U(1) gauge fields provided evidence of a massless photon in the theory's unconfined phase without having a single-photon operator, relying instead on an operator for a pair of photons.
Here we have constructed three different operators with single-photon transformation properties
$T_1^{--}$  and used them to confirm the massless photon.
The pattern of cross correlations among these operators shows that they are all
propagating the same photon state.
Our ``solenoidal'' operator (here called $O_4$) was found to give a particularly robust signal for the photon.

As a further application of the single photon operators we consider the fate of the 
photon in the abelian Higgs model.
Past lattice studies of this model have used a composite vector operator (here called $O_5$) built from both the scalar and gauge fields to expose the presence of a massive 
vector particle in the Higgs regime. Now that we have $T_1^{--}$ operators built purely from gauge fields ($O_2$, $O_3$ and $O_4$), cross correlations can be studied
and it was shown that the photon and composite vector operators couple to the 
same massive vector state.
It was also observed that there is some excited state contribution present in the correlator 
of $O_5$
which is absent from $O_2$, $O_3$ and $O_4$.
It may be interesting for future work to study these excited state effects in detail.

The photon operators discussed in this work can be applied in other field theories. 
Of particular relevance may be SU(2)$\times$U(1) gauge theory with a complex Higgs doublet, where the photon, customarily understood to emerge as a linear combination 
of the SU(2) and U(1) gauge fields, remains massless in the Higgs phase. The operators
discussed here could be useful for investigating the lattice version of this scenario.

\acknowledgments
The work of R.L. is supported in part by the Natural Sciences and 
Engineering Research Council of Canada.
TRIUMF receives federal funding via a contribution agreement 
with the National Research Council of Canada.

\appendix*
\section{Explicit operators}

The operators are constructed in a spatial lattice at a fixed Euclidean time.
Using $i,j,k$ to denote the 3 distinct spatial directions, the operators are

\begin{widetext}
\begin{eqnarray}
O_1^i(x) &=& {\rm Im}\left(P_{jk}(x)+P_{jk}(x-\hat j)
             +P_{jk}(x-\hat j-\hat k)+P_{jk}(x-\hat k)\right) \,, \\
O_2^i(x) &=& {\rm Im}\left(Q_{ij}(x)+Q_{ij}(x-\hat i)+Q_{ik}(x)+Q_{ik}(x-\hat i)\right) \,, \\
O_3^i(x) &=& {\rm Im}\left(R_{ij}(x)+S_{ij}(x)+R_{ik}(x)+S_{ik}(x)\right) \,, \\
O_4^i(x) &=& {\rm Im}\left(P_{ij}(x)+P_{ji}(x-\hat j)+P_{ij}(x-\hat i)+P_{ji}(x-\hat i-\hat j)
                           P_{ik}(x)+P_{ki}(x-\hat k)+P_{ik}(x-\hat i)+P_{ki}(x-\hat i-\hat k)\right) \,,
\nonumber \\
\end{eqnarray}
where
\begin{eqnarray}
P_{ij}(x) &=& U_i(x)U_j(x+\hat i)U_i^\dagger(x+\hat j)U_j^\dagger(x) \,, \\
Q_{ij}(x) &=& U_j(x)U_i(x+\hat j)U_j^\dagger(x+\hat i)U_i^\dagger(x)
              U_j^\dagger(x-\hat j)U_i(x-\hat j)U_j(x+\hat i-\hat j)U_i^\dagger(x) \,, \\
R_{ij}(x) &=& U_i(x)U_j(x+\hat i)U_i^\dagger(x+\hat j)U_j^\dagger(x)
              U_j^\dagger(x-\hat j)U_i^\dagger(x-\hat i-\hat j)U_j(x-\hat i-\hat j)U_i(x-\hat i) \,, \\
S_{ij}(x) &=& U_i(x)U_j^\dagger(x+\hat i-\hat j)U_i^\dagger(x-\hat j)U_j(x-\hat j)
              U_j(x)U_i^\dagger(x-\hat i+\hat j)U_j^\dagger(x-\hat i)U_i(x-\hat i) \,.
\end{eqnarray}
Another option for operator $O_2$ is given by
\begin{eqnarray}
O_2^i(x) &=& {\rm Im}\bigg(Q_{ij}(x)+Q_{ij}(x-\hat i)+Q_{ij}(x-\hat j)+Q_{ij}(x-\hat i-\hat j)
+Q_{ij}(x+\hat j)+Q_{ij}(x-\hat i+\hat j) \nonumber \\
&& +Q_{ik}(x)+Q_{ik}(x-\hat i)
+Q_{ik}(x-\hat k)+Q_{ik}(x-\hat i-\hat k)+Q_{ik}(x+\hat k)+Q_{ik}(x-\hat i+\hat k)\bigg) \,.
\end{eqnarray}
\end{widetext}

%%%%%%%%%%%%%%%%%%%%%%%%%%%

\end{document}